\documentstyle[psfig,draft]{mn}

\def\gs{\mathrel{\raise0.35ex\hbox{$\scriptstyle >$}\kern-0.6em
\lower0.40ex\hbox{{$\scriptstyle \sim$}}}}
\def\ls{\mathrel{\raise0.35ex\hbox{$\scriptstyle <$}\kern-0.6em
\lower0.40ex\hbox{{$\scriptstyle \sim$}}}}
\def\mnras {{MNRAS}}
\def\apj {ApJ}
\def\apjs {ApJS}

\def\aj {AJ}

\def\mpc {${\rm h^{-1}}$ Mpc}
\def\nat {Nature}

\title[Large-scale wall in Abell 22]
      {Discovery of a Large-scale Wall in the Direction of Abell 22}

\author[K.\,A.\ Pimbblet, A.\,C.\ Edge \& W.\,J.\ Couch]
       {Kevin A.\ Pimbblet,$^{\! 1,2}$ 
        Alastair C.\ Edge,$^{\! 3}$ 
        Warrick J.\ Couch $^{\! 4}$
        \vspace*{1mm}\\
 	$^1$ pimbblet@physics.uq.edu.au\\
	$^2$ Department of Physics, University of Queensland, Brisbane,
	QLD 4072, Australia\\
        $^3$ Department of Physics, University of Durham, South Road,
        Durham, DH1 3LE, UK\\
	$^4$ School of Physics, University of New South Wales, Sydney, 
	NSW 2052, Australia}

\date{\fbox{\sc Draft: \today\ --- Do Not Distribute}}

\pagerange{000--000}

\begin{document}

\maketitle

\begin{abstract}
We report on the discovery of a large-scale wall in the 
direction of Abell~22.
Using photometric and spectroscopic data 
from the Las Campanas Observatory and Anglo--Australian Telescope 
Rich Cluster Survey, Abell~22 is found to exhibit a 
highly unusual and striking redshift distribution.  
We show that Abell~22 exhibits a foreground wall-like structure by examining 
the galaxy distributions in both redshift space and on the colour-magnitude
plane.  A search for other galaxies and clusters in the nearby
region using the 2dF Galaxy Redshift Survey database
suggests that the wall-like structure is a significant large-scale,
non-virialized filament which runs 
between two other Abell clusters either side of Abell~22.  
The filament stretches over at least $>40$ \mpc \ 
in length and 10 \mpc \ in width at the redshift of Abell~22.

\end{abstract}

\begin{keywords}
galaxies: clusters: individual: Abell 22 -- 
surveys -- 
large-scale structure of Universe
\end{keywords}

\section{Introduction}

Hierarchical structure formation modelling
(Zeldovich, Einasto \& Shandarin 1982; Katz et al.\ 1996; 
Bond, Kofmann \& Pogosyan 1996; Jenkins et al.\ 1998; Colberg et al.\ 2000)
succinctly demonstrates how large scale structure such as
galaxy clusters grow from the non-isotropic infall and
accretion of smaller components.
Material infalling toward galaxy clusters is found to be 
preferentially situated along galaxy filaments (Bond, Kofmann \& 
Pogosyan 1996).  These filaments are, in turn, observed to
stretch between galaxy clusters
at all redshifts (e.g.\ Ebeling, Barrett, \& Donovan 2004;
Pimbblet \& Drinkwater 2004; 
Colberg, Krughoff \& Connolly 2004;
Pimbblet, Drinkwater \& Hawkrigg 2004; Dietrich et al.\ 2004;
Gal \& Lubin 2004; Durret et al.\ 2003;
Scharf et al.\ 2000; Doroshkevich et al.\ 2000; Colberg et al.\ 1999; 
Geller et al.\ 1997; Bond, Kofman \& Pogosyan 1996;
Shectman et al.\ 1996; Geller \& Huchra 1989)
and are also observed near very isolated clusters (Durret et al.\ 2003;
Kodama et al.\ 2001).
Since galaxy filaments can comprise over 40 per cent of the total 
cluster mass at clustocentric radii of 4--6.5 \mpc \ 
(Colberg et al.\ 1999; see also Cen \& Ostriker 1999)
their direct observation remains paramount to our
understanding of cluster growth.

Significantly, not all galaxy filaments are the same.  They exist in
many morphologies and varied lengths (Colberg, Krughoff \& Connolly 2004;
CKC herein;
Pimbblet, Drinkwater \& Hawkrigg 2004; PDH herein).  Shorter filaments
are generally found between close cluster pairs; longer filaments
are much rarer (e.g.\ straight filaments of 30 \mpc\footnote{
We use $H_0 = 100 \ 
h$ km s$^{-1}$\,Mpc$^{-1}$ and $q_o=0.5$ throughout this work.  
Further, all quoted coordinates are J2000 compliant.} 
in length are
at least 5 times as rare as those of 10 \mpc \ in length; PDH).
In the nomenclature of PDH, Type~III inter-cluster filament
morphologies (`walls' or `sheets' of galaxies) are exceedingly rare,
comprising \emph{at most} no more than 3 per cent of the total filament 
population (CKC; PDH).  In this work, we report on the
discovery of a large-scale ($>40$ \mpc) wall-like filament in the
direction of Abell~22 (X-ray centre is at
$\alpha = $ 00 20 38, $\delta = $ -25 43 19; 
$cz = 42676 \pm 98$ kms$^{\rm{-1}}$; Pimbblet 2001) found in the
Las Campanas Observatory and Anglo--Australian Telescope 
Rich Cluster Survey (LARCS; e.g.\ Pimbblet et al.\ 2001).
Briefly,
LARCS is a long-term project to study a statistically-reliable sample
of 21 of the most luminous X-ray clusters at intermediate redshifts
($0.07<z<0.16$) in the southern hemisphere.  We are mapping the
photometric, spectroscopic and dynamical properties of galaxies in rich
cluster environments at $z\sim 0.1$, tracing the variation in these
properties from the high-density cluster cores out into the surrounding
low-density field beyond the turn-around radius.  For the most massive
clusters at $z\sim 0.1$, the turn-around radius corresponds to roughly
1\,degree or a 10 \mpc \ radius (O'Hely et al.\ 1998; Pimbblet 2001) and 
therefore we have obtained panoramic CCD imaging covering
2-degree diameter fields, as well as spectroscopic coverage of these
fields (e.g. Pimbblet et al.\ 2001; O'Hely 2000).
The imaging comes from $B$ and $R$-band mosaics taken with the 1-m Swope
telescope at Las Campanas Observatory, while the spectroscopy comes from
the subsequent follow-up with the 400-fibre 2dF multi-object spectrograph
on the 3.9-m Anglo-Australian Telescope (AAT).

The plan of this paper is as follows:
In \S2 we examine the unusual velocity structure of Abell~22
and the foreground wall-like structure.
We then divide our observations into redshift bins 
and examine the distribution of the constituent galaxies on the 
colour-magnitude plane in \S3.  In \S4 we present 
a search for other local clusters to determine if the wall-like
structure is a part of a larger-scale object.  
We summarize our findings in \S5.

\section{Velocity and Spatial Structure}

The photometric data reduction for LARCS is presented in 
Pimbblet et al.\ (2001) whilst the spectroscopic 
reduction will be presented in Pimbblet et al.\ (in prep.).

\begin{figure}
\centerline{\psfig{file=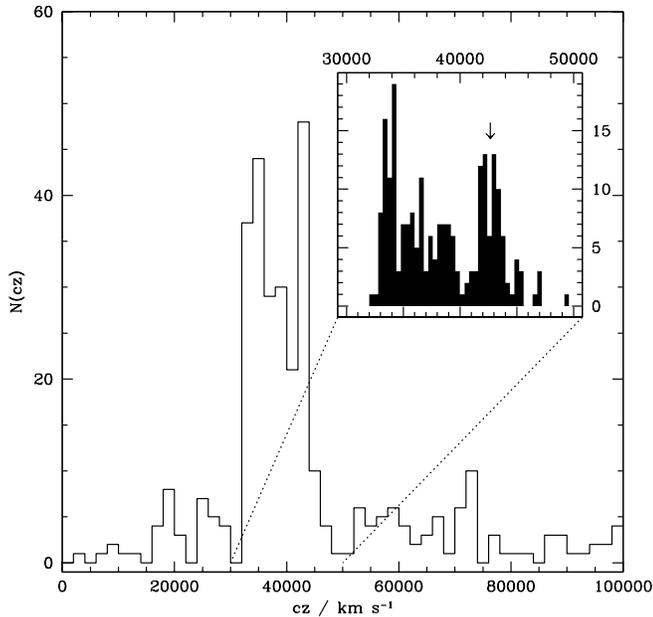,angle=0,width=3.5in}}
  \caption{\small{Velocity histogram for Abell~22 with inset
showing a magnification of the central region.
The bins are 1000 ${\rm km s^{-1}}$ wide in the main plot and 
400 ${\rm km s^{-1}}$ wide in the inset panel.  The downward arrow
denotes the velocity of Abell~22 as computed by Pimbblet (2001).
There is a lot of significant structure in the foreground of Abell~22.
}}
  \label{fig:velhist}
\end{figure}

\begin{figure}
\vspace*{-1.in}
\centerline{\psfig{file=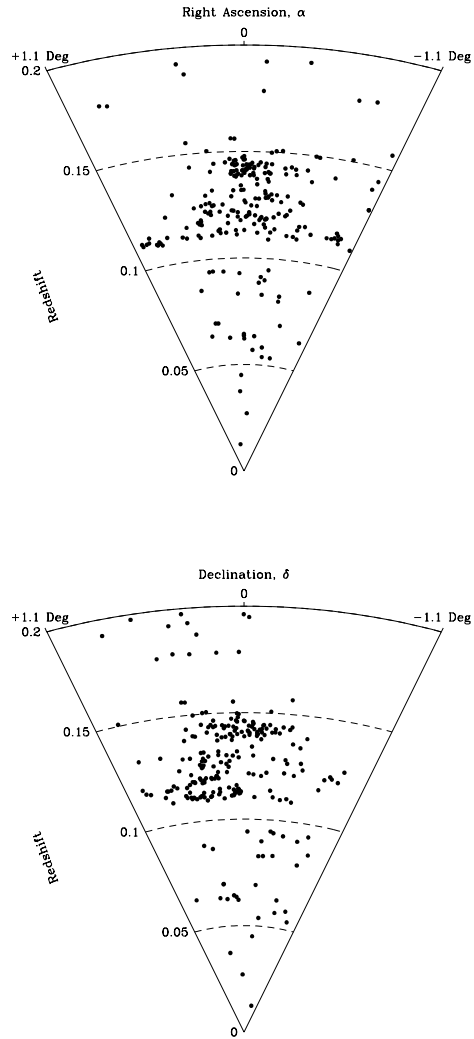,angle=0,width=5in,height=7.5in}}
\vspace*{-1.in}
  \caption{\small{Wedge plots of right ascension and declination 
versus redshift in the direction of Abell~22.  The X-ray centre of the
cluster is located at $\alpha = \delta = 0$.  Note the ensemble of
galaxies in the foreground of the cluster.
Although one may expect this plot to exhibit a `finger of God'
effect for the cluster, arising from the distorting effects in redshift
space, this is not seen as the elongation in redshift space of 
the cluster is small in comparison to the depth of redshift 
space covered.
}}
  \label{fig:wplot22}
\end{figure}

Observations of Abell~22 using 2dF (Pimbblet 2001) have 
yielded redshifts for 345 galaxies. 
From these, it is possible to see the presence of an unusual structure
from the redshift distribution, shown in Figure~\ref{fig:velhist}. 
Figure~\ref{fig:wplot22} shows the position of the galaxies
along the line of sight with respect to their angular position
on the sky.

%
%

%
%

Abell~22 has a central peak which coincides with \emph{some} previously 
published estimates of the clusters redshift (e.g. Dalton et al.\ 1994).
The peak of the distribution in the velocity histogram is 
unusually broad and contains multiple sub-peaks.  
The wedge plot (Figure~\ref{fig:wplot22}) reveals that Abell~22
is a highly complex structure.    
The foreground galaxies are striking: a non-virialized wall-like structure 
appears to stretch across the entire RA range of the observations,
but is more limited in scope (and offset from the cluster) in Dec.
It is little wonder, therefore, that other redshift
estimates for Abell~22 differ from that of Dalton et al.\ (1994) 
by as much as $\sim 3000 {\rm km s^{-1}}$ (e.g. Struble \& Rood, 1999).

\section{The Colour-Magnitude Relation}

%
%
\begin{table}
\begin{center}
\medskip
\caption{\small{Fraction of blue galaxies, $f_B$, within each redshift
bin; where $f_B$ is defined here as being the fraction brighter than
$M^{*}+1$ possessing a rest frame colour $\Delta(B-V)=-0.2$ bluer than 
the fitted CMR of Pimbblet et al.\ (2002) over all galaxies brighter 
than $M^{*}+1$.  The column headed $N(M<M^{*}+1)$ is the number of
galaxies brighter than $M^{*}+1$.
}}
\begin{tabular}{lcc}
\hline
Redshift Bin & $N(M<M^{*}+1)$	&  Blue Fraction,\\
(${\rm{kms^{-1}}}$) & 			&  $f_B$ \\
\hline 
43126-45000	& 19 & $0.05 \pm 0.05$ \\
41251-43125	& 33 & $0.18 \pm 0.07$ \\
39376-41250	& ~9 & $0.22 \pm 0.16$ \\
37501-39375	& 21 & $0.29 \pm 0.12$ \\
35626-37500	& 25 & $0.52 \pm 0.14$ \\
33751-35625	& 32 & $0.31 \pm 0.10$ \\
31876-33750	& 26 & $0.35 \pm 0.12$ \\
\hline
\end{tabular}
\label{tab:fb}
\end{center}
\end{table}

\begin{figure}
\centerline{\psfig{file=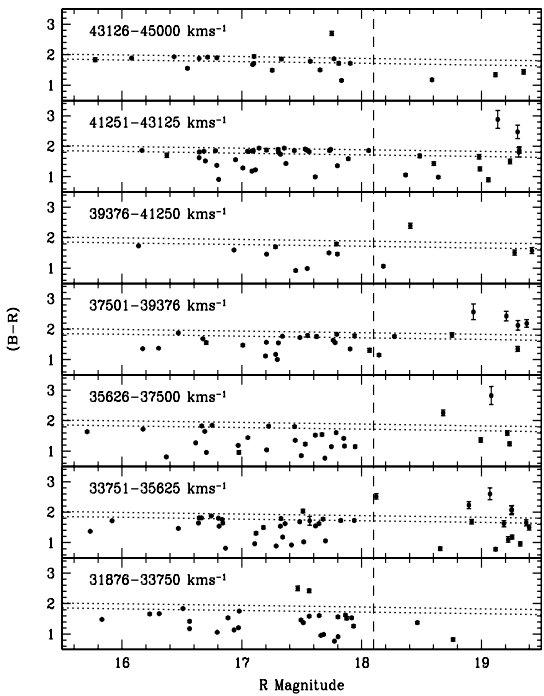,angle=0,width=3.5in,height=4in}}
  \caption{\small{Colour-magnitude diagrams for each of the seven
redshift bins.  The vertical dashed line denotes the $M^{*}+1$
magnitude division used during the 2dF observations (Pimbblet 2001).  
The dotted lines are the 1$\sigma$ error to the inner ($r_p<1$Mpc) 
region CMR fit from Pimbblet et al.\ (2002). 
}}
  \label{fig:cmr}
\end{figure}

We now split our sample up into seven equal 
redshift bins from 31875 ${\rm{kms^{-1}}}$ to 45000 ${\rm{kms^{-1}}}$
and construct colour-magnitude diagrams for each bin.
These diagrams are display in Figure~\ref{fig:cmr}.

Pimbblet et al.\ (2002) investigated the environmental dependence
of galaxy colours within a sample of eleven LARCS clusters; including
Abell~22.
They construct colour-magnitude diagrams divided into radial
intervals for these clusters and fitted the colour-magnitude relation
(CMR; e.g. Visvanathan \& Sandage 1977) using a biweight method.
Here, we use the errors on their CMR fit to Abell~22 as a \emph{guideline}
for colour magnitude diagrams constructed for each of the redshift
bins (Figure~\ref{fig:cmr}).

%
%

The CMR distribution for the redshift bins nearer to the 
filament are qualitatively similar to those nearer to the cluster
(Figure~\ref{fig:cmr}).  
To quantify the similarity of the distributions, 
we derive the blue fraction, $f_B$, of
galaxies in each redshift bin.  Here, we define $f_B$ to be the
fraction of galaxies brighter than $M^{*}+1$ possessing a
rest frame colour that is $\Delta(B-V)=-0.2$ bluer than the fitted 
CMR of Pimbblet et al.\ (2002).  This differs from the classical 
Butcher-Oemler (1984; BO; see Pimbblet 2003 for a review) 
definition of $f_B$ 
(i.e. our $M^{*}+1$ magnitude cut does not correspond to the $M_V=-20$ 
magnitude cut used by BO and we have not applied a limiting 
clustocentric radius),
but this is a sound approach as we are simply interested in 
internal differences in $f_B$ between our redshift bins.
Table~\ref{tab:fb} lists the blue fractions for the redshift bins
together with associated Poissonian errors.  
The blue fraction increases from the upper two redshift bins
(i.e. Abell~22) to the lower redshift bin (i.e. the wall-like 
structure), but only by 1--2$\sigma$.  Given the large errors on these
numbers and the low number of galaxies in some bins (see 
Table~\ref{tab:fb}) we do not view this as a highly significant change.
Moreover, the blue fraction of the filament is very comparable to that 
found by Pimbblet \& Drinkwater (2004).

\section{Discussion}

To determine whether the wall-like structure displayed 
in Figure~\ref{fig:wplot22} may be a part of a more
extended object we utilize the NASA/IPAC Extragalactic 
Database (NED) to look for any near-by (within 3 degrees on the sky)
clusters of galaxies at similar redshifts.
Four clusters in addition to Abell~22 are found and we list all five
of these in Table~\ref{tab:other}.
Given the redshifts of Abell~14 and APMCC~029, it is unlikely that
they are associated with the filament.  Meanwhile, Abell~22 and Abell~47
certainly reside in the background of the filament and are likely 
unconnected due to the huge physical scale between them and
the filament.
We note that Abell~15, however, resides at a redshift that is 
similar to the filament, but is Abell~15 connected 
to the filament?

%
%
\begin{table*}
\begin{center}
\medskip
\caption{\small{Position and redshifts of all clusters with measured 
redshifts within three degrees of Abell~22 found using NED.
The column headed `Redshift Reference' gives the source of the quoted
redshift. 
}}
\begin{tabular}{lccccl}
\hline
Cluster  &  RA  &  Dec  & Distance from & Redshift & Redshift \\
Name & \multispan2{\hfil (J2000) \hfil } & Abell~22 ($'$) & (${\rm{kms^{-1}}}$) & Reference \\
\hline 
Abell~22$^\ddagger$ & 00 20 34.8 & -25 41 51 & n/a & 42658 & Pimbblet (2001) \\
Abell~15 & 00 15 13.7 & -26 01 19 & 75 & 36043 & De Propris et al.\ (2002) \\
APMCC~029 & 00 15 04.0 & -24 00 04 & 126 & 18887 & Dalton et al.\ (1997) \\
Abell~14 & 00 15 13.9 & -23 53 19 & 131 & 19636 & Struble \& Rood (1999) \\
Abell~47 & 00 30 35.9 & -24 09 26 & 165 & 41431 & Struble \& Rood (1999) \\
\hline
\end{tabular}

$^\ddagger$ Coordinates for Abell~22 presented here are from NED.  They
differ slightly to the coordinates \\
of the cluster's X-ray centre as noted in section~1.

\label{tab:other}
\end{center}
\end{table*}

\begin{figure*}
\centerline{\psfig{file=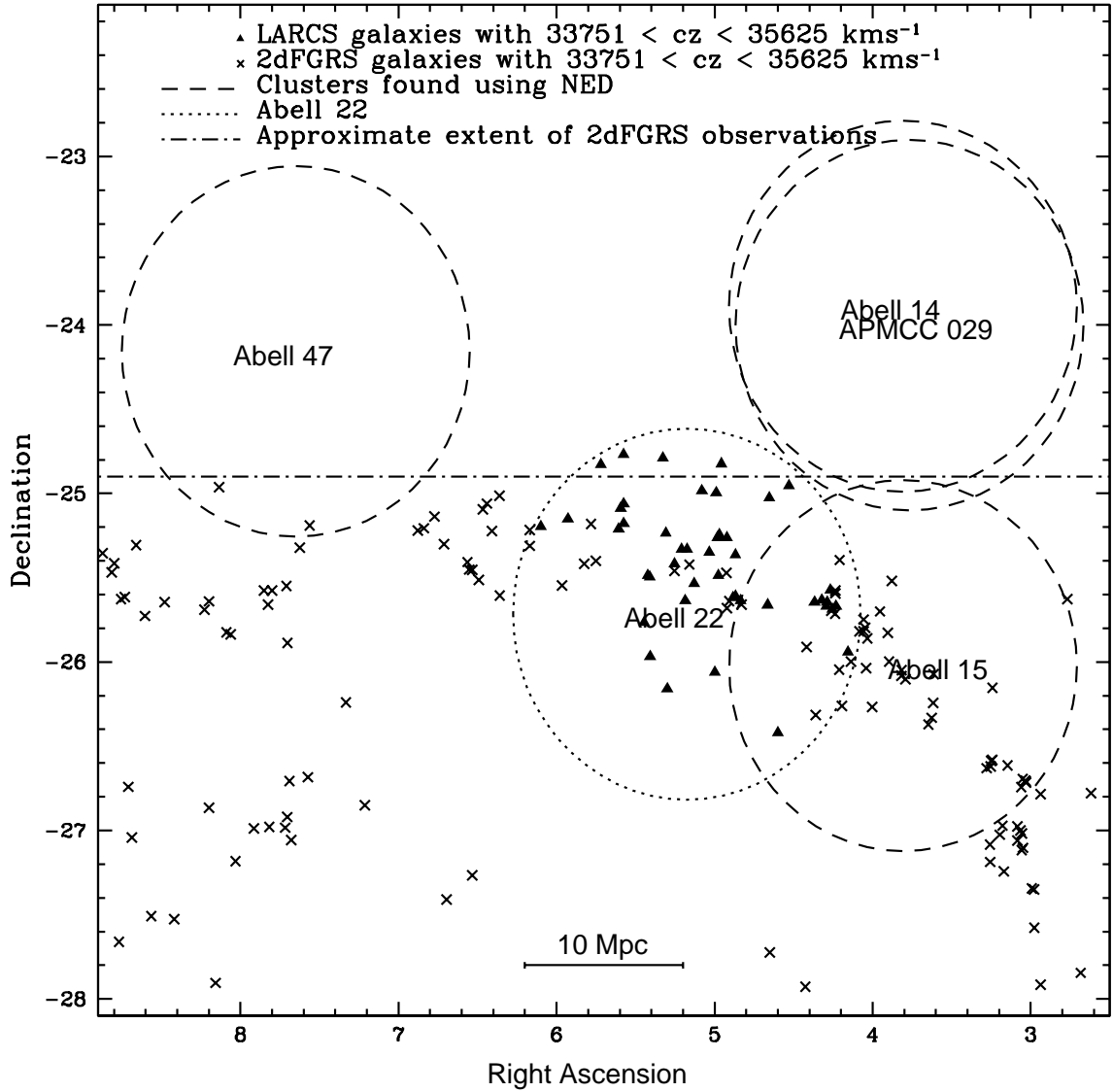,angle=0,width=6.5in,height=6.5in}}
  \caption{\small{Spatial distribution of galaxies contained
in the wall-like structure (filled triangles),
galaxies contained in the 2dFGRS final data release (crosses)
and all clusters with measured redshifts 
within 3 degrees of Abell~22 (large circles of diameter
2 degrees).
Note that the 2dFGRS final data release does not
extend across all of this area; its approximate extent is 
denoted by the dot-dashed line.
The scale bar shows 10 \mpc \ at the central redshift of the 33751--35625
${\rm{kms^{-1}}}$ bin.
We suggest the wall-like structure extends from beyond Abell~15, across
Abell~22 and toward Abell~47.  
}}
  \label{fig:other}
\end{figure*}

Since this region overlaps (Pimbblet et al.\ 2001) with the
2dF Galaxy Redshift Survey (2dFGRS; Colless et al.\ 2001)
we now use the final release of the 2dFGRS 
(see http://www.mso.anu.edu.au/2dFGRS/)
to search for any galaxies
with measured redshifts in the range 33751--35625 ${\rm{kms^{-1}}}$, 
thereby encompassing our final redshift bin.
The result of both of these searches are illustrated 
in Figure~\ref{fig:other}.
We note that the 2dFGRS final data release 
does not extend above $\delta\sim-24.9$ but does encompass
Abell~15, much of Abell~22 and partial amounts of Abell~47 
(Figure~\ref{fig:other}).  
We can therefore use 2dFGRS to try to answer the question of 
whether Abell~15 is connected to the filament. 
Since the velocity dispersion for
Abell~15 is $\sigma_z \sim500$ kms$^{\rm{-1}}$\footnote{De 
Propris et al.\ (2002)
note that their observations for Abell~15 are 54 per cent
complete ($N_{gal}=24$), hence we use a velocity dispersion
of $\sim500$ kms$^{\rm{-1}}$ as an order of magnitude
estimate only.  The actual value calculated by them is 
$\sigma_z = 459^{+104}_{-82}$ kms$^{\rm{-1}}$.} 
(De Propris et al.\ 2002), the filament easily resides within
it's infall regions (Diaferio \& Geller 1997)
at $2 < \Delta v / \sigma_z < 4$ (see Carlberg, Yee \& Ellingson 1997). 
An inspection of galaxy surface density contours around Abell~15 
and Abell~47 also 
shows that these clusters are elongated in the direction of 
the filament (Figure~\ref{fig:other}; see Binggeli 1982;
Plionis et al.\ 2003 and references therein) meaning they are
very likely to be connected. 

%
%

We suggest that the wall-like structure
is a significant non-virialized overdensity in galaxies
which runs from (at least) Abell~15, across Abell~22 and toward
Abell~47.  If so, then the filament stretches over at least $>40$ 
\mpc \ at this redshift.  
The surface density (computed in the same way 
as Pimbblet et al.\ 2002) of the filament 
is typically $\sim$10 galaxies per square \mpc.
Such a surface density is very similar to that found in other 
filaments (Ebeling, Barrett \& Donovan 2004; PDH).
As the filament connects two clusters, we cannot yet consider 
it to be of the same ilk as the Great Wall 
noted by Geller \& Huchra (1989).  It may however, be taken as evidence
that the region around Abell~22 is a supercluster candidate.

\section{Summary}

We have presented evidence for a new large-scale wall in the
direction of Abell~22.  

\smallskip

\begin{itemize}

\item  The velocity structure of Abell~22 is highly unusual:
the cluster appears to extend in
redshift space and merges with foreground galaxy structure.   
These foreground galaxies are highly striking: a wall-like structure 
appears to stretch across the entire RA range of the observations.
It is probable that the presence of this wall-like structure
has caused previous estimates of Abell~22's redshift to differ
significantly.

\item  Colour-magnitude diagrams for the redshift bins 
are qualitatively similar and the blue fraction of galaxies does not
change significantly between the cluster and the foreground filament.

\item  The wall appears to be part of a large-scale filament
that passes in front of Abell~22 and Abell~47, and is likely 
connected to Abell~15, which is at a similar redshift
to the filament and elongated toward it.
At this redshift, the filament stretches over at least 40 \mpc.
We suggest that this region is a supercluster candidate.
\end{itemize}

Further spectroscopic observations in the direction of 
Abell~47 are required to definitively determine the filaments extension
toward this cluster and to find out if the structure is part
of a larger supercluster.  
It would also be interesting to investigate if the blue fraction
and morphology of galaxies along other filaments
(e.g.\ PDH) remained approximately constant along their length
and how similar is it to clusters they connect to?  
Indeed the morphology-density relation (Dressler 1980) of filaments 
remains a relatively unstudied area.

\subsection*{Acknowledgements}

We thank the anonymous referee for a very speedy and useful review.
We also thank Mary Hawkrigg and Nathan Courtney for reading through 
an earlier draft of this work and providing useful comments.
KAP acknowledges support from an EPSA University of Queensland
Research Fellowship and a UQRSF grant.  
ACE acknowledges support from the Royal Society.  
WJC acknowledges the financial support of the Australian 
Research Council.
We thank the Observatories of the Carnegie Institution of Washington 
and the Anglo-Australian Observatory for their
generous support of this survey.  
In particular, we warmly thank Geraint Lewis and 
Terry Bridges for their assistance at the telescope.
 
This research has made use of the NASA/IPAC Extragalactic Database 
(NED) which is operated by the Jet Propulsion Laboratory, California 
Institute of Technology, under contract with the National Aeronautics 
and Space Administration.


\begin{thebibliography}{}

\bibitem[\protect\citeauthoryear{Binggeli}{1982}]{1982A&A...107..338B} 
Binggeli B., 1982, A\&A, 107, 338 

\bibitem[\protect\citeauthoryear{Bond et al.}{1996}]{1996bond}
Bond J.~R., Kofman L., Pogosyan D., 1996, \nat, 380, 603

\bibitem[Butcher \& Oemler (1984)]{1984ApJ...285..426B} Butcher, H.\ \& Oemler, A.\ 1984, \apj, 285, 426 (BO)

\bibitem[\protect\citeauthoryear{Carlberg, Yee, \& 
Ellingson}{1997}]{1997ApJ...478..462C} Carlberg R.~G., Yee H.~K.~C., 
Ellingson E., 1997, ApJ, 478, 462 

\bibitem[\protect\citeauthoryear{Cen \& 
Ostriker}{1999}]{1999ApJ...514....1C} Cen R., Ostriker J.~P., 1999, ApJ, 
514, 1 

\bibitem[\protect\citeauthoryear{Colberg, Krughoff, \& 
Connolly}{2004}]{2004astro.ph..6665C} Colberg J.~M., Krughoff K.~S., 
Connolly A.~J., 2004, astro, astro-ph/0406665 (CKC)

\bibitem[\protect\citeauthoryear{Colberg et 
al.}{2000}]{2000MNRAS.319..209C} Colberg J.~M., et al., 2000, MNRAS, 319, 
209 

\bibitem[\protect\citeauthoryear{Colberg et 
al.}{1999}]{1999MNRAS.308..593C} Colberg J.~M., White S.~D.~M., Jenkins A., 
Pearce F.~R., 1999, MNRAS, 308, 593 

\bibitem[\protect\citeauthoryear{Colless et 
al.}{2001}]{2001MNRAS.328.1039C} Colless M.~et al., 2001, MNRAS, 328, 1039 

\bibitem[Dalton, Maddox, Sutherland, \& 
Efstathiou(1997)]{1997MNRAS.289..263D} Dalton, G.~B., Maddox, S.~J., 
Sutherland, W.~J., \& Efstathiou, G., 1997, \mnras, 289, 263 

\bibitem[Dalton, Efstathiou, Maddox, \& Sutherland(1994)]{1994MNRAS.269..151D} 
Dalton, G.~B., Efstathiou, G., Maddox, S.~J., \& Sutherland, W.~J., 1994, 
\mnras, 269, 151 

\bibitem[\protect\citeauthoryear{De Propris et 
al.}{2002}]{2002MNRAS.329...87D} De Propris R., et al., 2002, MNRAS, 329, 
87 

\bibitem[\protect\citeauthoryear{Diaferio \& 
Geller}{1997}]{1997ApJ...481..633D} Diaferio A., Geller M.~J., 1997, ApJ, 
481, 633 

\bibitem[\protect\citeauthoryear{Dietrich et 
al.}{2004}]{2004astro.ph..6541D} Dietrich J.~P., Schneider P., Clowe D., 
Romano-Diaz E., Kerp J., 2004, astro, astro-ph/0406541 

\bibitem[Doroshkevich et al.(2000)]{2000MNRAS.315..767D} Doroshkevich, 
A.~G., Fong, R., McCracken, H.~J., Ratcliffe, A., Shanks, T., \& 
Turchaninov, V.~I.\ 2000, \mnras, 315, 767 

\bibitem[\protect\citeauthoryear{Dressler}{1980}]{1980ApJ...236..351D} 
Dressler A., 1980, ApJ, 236, 351 

\bibitem[\protect\citeauthoryear{Durret et al.}{2003}]{2003A&A...403L..29D} 
Durret F., Lima Neto G.~B., Forman W., Churazov E., 2003, A\&A, 403, L29 

\bibitem[\protect\citeauthoryear{Ebeling, Barrett, \& 
Donovan}{2004}]{2004ApJ...609L..49E} Ebeling H., Barrett E., Donovan D., 
2004, ApJ, 609, L49 

\bibitem[\protect\citeauthoryear{Gal \& Lubin}{2004}]{2004ApJ...607L...1G} 
Gal R.~R., Lubin L.~M., 2004, ApJ, 607, L1 

\bibitem[Geller \& Huchra(1989)]{1989Sci...246..897G} Geller, M.~J.~\& 
Huchra, J.~P.\ 1989, Science, 246, 897 

\bibitem[Geller et al.(1997)]{1997AJ....114.2205G} Geller, M.~J.~et al.\ 
1997, \aj, 114, 2205 

\bibitem[\protect\citename{Jenkins} 1998]{1998ApJ...499...20J} Jenkins A., 
Frenk C.~S., Pearce F.~R., et al., 1998, ApJ,  499, 20 

\bibitem[\protect\citename{Katz} 1996]{1996ApJ...457L..57K} Katz N., 
Weinberg D.~H., Hernquist L., Miralda-Escude J., 1996, ApJ,  457, L57 

\bibitem[\protect\citeauthoryear{Kodama et al.}{2001}]{2001ApJ...562L...9K} 
Kodama T., Smail I., Nakata F., Okamura S., Bower R.~G., 2001, ApJ, 562, L9 

\bibitem{10b} O'Hely, E., 2000, Ph.D. Thesis, University of New South Wales

\bibitem{10} O'Hely, E., Couch, W.J., Smail, I., Edge, A.C., Zabludoff, A.\ I., 1998, PASA, 15, 273

\bibitem[Pimbblet (2001)]{2001P} Pimbblet, K.~A, 2001, Ph.D. Thesis, University of Durham

\bibitem[\protect\citeauthoryear{Pimbblet et 
al.}{2001}]{2001MNRAS.327..588P} Pimbblet K.~A., Smail I., Edge A.~C., 
Couch W.~J., O'Hely E., Zabludoff A.~I., 2001, MNRAS, 327, 588 

\bibitem[\protect\citeauthoryear{Pimbblet et 
al.}{2002}]{2002MNRAS.331..333P} Pimbblet K.~A., Smail I., Kodama T., Couch 
W.~J., Edge A.~C., Zabludoff A.~I., O'Hely E., 2002, MNRAS, 331, 333 

\bibitem[\protect\citeauthoryear{Pimbblet}{2003}]{2003PASA...20..294P} 
Pimbblet K.~A., 2003, PASA, 20, 294 

\bibitem[\protect\citeauthoryear{Pimbblet \& 
Drinkwater}{2004}]{2004MNRAS.347..137P} Pimbblet K.~A., Drinkwater M.~J., 
2004, MNRAS, 347, 137 

\bibitem[\protect\citeauthoryear{Pimbblet, Drinkwater, \& 
Hawkrigg}{2004}]{2004MNRAS.354L..61P} Pimbblet K.~A., Drinkwater M.~J., 
Hawkrigg M.~C., 2004, MNRAS, 354, L61 (PDH)

\bibitem[\protect\citeauthoryear{Plionis et 
al.}{2003}]{2003ApJ...594..144P} Plionis M., Benoist C., Maurogordato S., 
Ferrari C., Basilakos S., 2003, ApJ, 594, 144 

\bibitem[\protect\citeauthoryear{Scharf et al.}{2000}]{2000ApJ...528L..73S} 
Scharf C., Donahue M., Voit G.~M., Rosati P., Postman M., 2000, ApJ, 528, 
L73 

\bibitem[Shectman et al.(1996)]{1996ApJ...470..172S} Shectman, S.~A., 
Landy, S.~D., Oemler, A., Tucker, D.~L., Lin, H., Kirshner, R.~P., \& 
Schechter, P.~L.\ 1996, \apj, 470, 172 

\bibitem[Struble \& Rood(1999)]{1999ApJS..125...35S} Struble, M.~F.~\& 
Rood, H.~J., 1999, \apjs, 125, 35 

\bibitem[\protect\citeauthoryear{Visvanathan \& 
Sandage}{1977}]{1977ApJ...216..214V} Visvanathan N., Sandage A., 1977, ApJ, 
216, 214 

\bibitem[\protect\citename{Zeldovich} 1982]{1982Natur.300..407Z} Zeldovich 
I.~B., Einasto J., Shandarin S.~F., 1982, \nat,  300, 407 



\end{thebibliography}
\end{document}